\documentclass[pdflatex,sn-mathphys-num]{sn-jnl}%

\usepackage{graphicx}%
\usepackage{multirow}%
\usepackage{amsmath,amssymb,amsfonts}%
\usepackage{amsthm}%
\usepackage{mathrsfs}%
\usepackage[title]{appendix}%
\usepackage{xcolor}%
\usepackage{textcomp}%
\usepackage{manyfoot}%
\usepackage{booktabs}%
\usepackage{array}
\newlength{\pwtablewidth}
\usepackage{algorithm}%
\usepackage{algorithmicx}%
\usepackage{algpseudocode}%
\usepackage{listings}%
\usepackage{cleveref}

\theoremstyle{thmstyleone}%

\theoremstyle{thmstyletwo}%
\newtheorem{remark}{Remark}%

\theoremstyle{thmstylethree}%

\DeclareMathOperator*{\argmin}{arg\,min}
\newcommand{\Nout}{N_{\text{out}}}
\newcommand{\Nin}{N_{\text{in}}}
\newcommand{\Ns}{N_s}
\newcommand{\tstart}{t_{\text{start}}}
\newcommand{\tmax}{t_{\text{max}}}
\newcommand{\A}{\mathcal{A}}
\newcommand{\DataUpdate}{\mathcal{G}_{n}^{\Phi}}

\newcommand{\AdamFirst}{\boldsymbol{\mu}}
\newcommand{\AdamSecond}{\boldsymbol{\nu}}
\newcommand{\dmAdam}{\Delta \mathbf{m}}

\begin{document}

\title{An exploration of plug-and-play flow matching for full-waveform inversion on breast imaging }

\author[1]{\fnm{Cristina} \sur{Garcia-Cardona}}\email{cgarciac@lanl.gov}

\author*[2]{\fnm{Siting} \sur{Liu}}\email{sitingl@ucr.edu}

\author[3]{\fnm{Fatma} \sur{Terzioglu}}\email{fterzioglu@ncsu.edu}

\author[4]{\fnm{Shanyin} \sur{Tong}}\email{tong3@sas.upenn.edu}

\author[5]{\fnm{Lu} \sur{Zhang}}\email{lz82@rice.edu}

\affil[1]{\orgdiv{Computing \& Artificial Intelligence Division}, \orgname{Los Alamos National Laboratory}, \orgaddress{\city{Los Alamos}, \postcode{87545}, \state{NM}, \country{USA}}}

\affil*[2]{\orgdiv{Department of Mathematics}, \orgname{University of California, Riverside}, \orgaddress{\city{Riverside}, \postcode{92521}, \state{CA}, \country{USA}}}

\affil[3]{\orgdiv{Department of Mathematics}, \orgname{North Carolina State University}, \orgaddress{\city{Raleigh}, \postcode{27695}, \state{NC}, \country{USA}}}

\affil[4]{\orgdiv{Department of Mathematics}, \orgname{University of Pennsylvania}, \orgaddress{\city{Philadelphia}, \postcode{19104}, \state{PA}, \country{USA}}}

\affil[5]{\orgdiv{Department of Computational Applied Mathematics and Operations Research}, \orgname{Rice University}, \orgaddress{\city{Houston}, \postcode{77005}, \state{TX}, \country{USA}}}

\abstract{
{Full-waveform inversion (FWI) reconstructs sound-speed fields from measured waveforms, by solving an optimization problem that minimizes the discrepancy between measured waveforms and those predicted by a numerical wave-propagation model. However, choosing suitable regularization for this optimization is challenging because it must stabilize the inversion while preserving relevant structures, prompting us to investigate whether priors learned from representative sound-speed images can improve reconstruction.
In this work, we adapt plug-and-play flow matching (PnP-Flow) to FWI by alternating waveform-misfit updates with updates from a pretrained flow prior. We explore the effectiveness of PnP-Flow reconstruction by comparing it with total-variation (TV) regularization using noiseless transmission-only plane-wave data.
Separate flow priors are trained on two circular-inclusion model families with constant and spatially varying backgrounds. We evaluate the method both on these in-distribution circle models and under substantial distribution shift by transferring the Gaussian-circle prior and reconstruction settings to breast phantom patches without retraining or retuning.
Using the same number of waveform-misfit updates, the proposed PnP-Flow framework generally improves circle reconstruction accuracy and recovers breast lesion contrast more accurately than TV baselines.
}
}


\maketitle

\section{Introduction}\label{sec:intro}
Full-waveform inversion (FWI) seeks to reconstruct spatially varying physical parameters governing a wave equation by solving an optimization problem that minimizes the discrepancy between measured waveforms and those predicted by a numerical wave-propagation model \cite{akcelik2002parallel,bunks1995multiscale,chen2020time}. Originally developed primarily in geophysical imaging \cite{tarantola1984inversion,virieux2010overview}, where it is used to infer subsurface material properties from seismic waveform measurements, FWI has increasingly been applied to other wave-based inverse problems, including medical ultrasound imaging \cite{bachmann2020source,bernard2017ultrasonic,wiskin2012nonlinear,sandhu2015frequency,lucka2022high}. In ultrasound tomography (UST), FWI is particularly promising because array-based acquisitions can provide measurements over many transmitter--receiver pairs and a broad range of frequencies, yielding rich waveform data for high-resolution quantitative reconstruction of tissue acoustic properties \cite{wang2015waveform,bachmann2020source}.

FWI has been studied extensively for quantitative breast UST. Early studies demonstrated the feasibility of waveform-based reconstruction of acoustic properties from breast ultrasound measurements
\cite{pratt2007sound,wiskin2012nonlinear}, while subsequent work improved computational efficiency and robustness through techniques such as source encoding, frequency continuation, data preprocessing, and improved initialization \cite{wang2015waveform,sandhu2015frequency}. Further developments incorporated attenuation modeling and explicit regularization \cite{perezliva2017time,matthews2017regularized}. More recently, computational advances have enabled high-resolution three-dimensional breast FWI \cite{lucka2022high}, as well as joint sound-speed and attenuation reconstruction from experimental in vitro and in vivo ring-array measurements \cite{ali2024slicewise}. Beyond breast UST, FWI has also been investigated for quantitative bone imaging \cite{bernard2017ultrasonic,fincke2022quantitative}
and transcranial ultrasound imaging, including high-resolution numerical studies, dedicated acquisition systems, and experimental brain-phantom reconstructions \cite{guasch2020full,cudeiro2022design,robins2023dual}. Recent applications to in vivo ultrasound measurements further demonstrate the growing potential of FWI in realistic medical-imaging settings \cite{ulrich2026quantitative}.

Despite this potential, FWI remains challenging. The parameter-to-data map is strongly nonlinear, and the resulting waveform-misfit objective is generally nonconvex. Gradient-based methods can therefore converge to undesirable local minima, particularly when corresponding features in the predicted and measured waveforms are misaligned by more than approximately half a cycle, leading to the phenomenon known as cycle skipping \cite{bunks1995multiscale,sirgue2004efficient,virieux2010overview}. FWI can consequently be highly sensitive to initialization. Moreover, finite acquisition coverage, limited frequency content, measurement noise, and modeling errors may leave some features of the unknown medium only weakly constrained by the data \cite{virieux2010overview}. These difficulties motivate the incorporation of prior information into the inversion.

A standard approach to incorporating prior information into FWI is to introduce explicit regularization terms that promote properties such as smoothness, sparsity, or piecewise regularity
\cite{matthews2017regularized}. Although such classical regularizers can substantially improve the stability of FWI, the prior information they encode is constrained by the prescribed form of the penalty. Regularizers based on smoothness or sparsity, for example, impose relatively simple structural assumptions on the unknown and may not fully reflect the statistical characteristics of a broader class of admissible wave-speed fields. Learned priors provide a more flexible alternative by extracting statistical and structural information directly from representative training data \cite{arridge2019solving}. Plug-and-play (PnP) methods provide one mechanism for incorporating such priors into inverse problems by separating the data-consistency step from the image-prior model \cite{venkatakrishnan-2013-pnp}. This framework has evolved from the use of classical denoisers such as BM3D \cite{dabov-2008-image} and learned convolutional denoisers such as DnCNN \cite{zhang-2017-dncnn} to increasingly expressive deep generative models, including score-based, diffusion, and flow matching models \cite{song-2019-score,ho-2020-ddpm,song-2021-sde,lipman-2023-cfm}. Several strategies have consequently been proposed for incorporating deep generative models into PnP-type inverse-problem solvers \cite{graikos-2022-diff_pnp,zhu-2023-ddm_pnp,wu-2024-diff_pnp,park-2024-scorepnp,martin2025pnpflow}.

Learning-based approaches have likewise begun to be incorporated into medical FWI. These include learned extrapolation of missing low-frequency data prior to conventional FWI \cite{robins2021deep}, decomposition-descent learning for transcranial ultrasound imaging \cite{tong2022transcranial}, deep unfolding of iterative reconstruction \cite{cohen2025deep}, and neural-network parameterizations or implicit regularizers for the unknown sound-speed field \cite{yan2025untrained,wang2025p2inr,cui2025fullwave}. Related developments outside medical imaging include untrained deep image priors \cite{wang2023rnn}, plug-and-play formulations based on classical or pretrained deep denoisers \cite{fu2023efficient,zhang2025progressive}, diffusion-based generative priors coupled with FWI \cite{wang2023diffusion,shan2026regularization,shen2026diffusion,zheng2025inversebench},
and
 hybrid approaches combining learning and wave physics \cite{borcea2026romnet,ding2022coupling}; see also the recent survey \cite{lin2026survey}.

Learned priors, however, can become less reliable when the target differs from the distribution represented in the training data.
In such cases, the learned update may favor features characteristic of the training ensemble while attenuating structures that are important for fitting the measured waveforms. This motivates examining the method not only in an in-distribution setting, but also under a substantial distribution shift.

Motivated by the PnP-Flow framework of \cite{martin2025pnpflow}, we develop a FWI reconstruction method that alternates blocks of waveform-misfit updates with updates induced by a pretrained flow-matching prior, with the final PnP iterate taken as the reconstruction. Our numerical study addresses two main questions: whether the proposed procedure improves small-inclusion reconstruction relative to TV-FWI using the same number of waveform-misfit updates, and whether the learned prior remains beneficial for recovering localized features under substantial distribution shift without additional tuning. To capture both global reconstruction quality and the recovery of small localized features, we evaluate whole-image reconstruction error together with lesion-contrast error, compared with total-variation-regularized FWI (TV-FWI) baseline.
Using the same number of waveform-misfit updates, PnP-Flow generally achieves lower reconstruction errors on circular-inclusion models and lower lesion-contrast errors on breast phantom images than TV-FWI.

The main contributions of this work are as follows:
\begin{itemize}
    \item
    We adapt the PnP-Flow framework to nonlinear full-waveform inversion by coupling a pretrained flow-matching prior with blocks of waveform-misfit updates. The prior is learned solely from sound-speed fields, whereas the data-consistency updates incorporate the wave-equation forward model and its adjoint.
    \item Using separately trained priors for two circular-inclusion model families, we show that the proposed method achieves lower reconstruction errors than the selected TV-FWI configurations under the same number of waveform-misfit updates.
    \item We assess out-of-distribution transfer by applying the Gaussian-circle prior and reconstruction settings to breast-phantom patches without retraining or retuning, and show improved lesion-contrast recovery relative to TV-FWI using correspondingly transferred settings.
\end{itemize}

The remainder of the paper is organized as follows. \Cref{sec:background} introduces the full-waveform inversion problem and discusses the challenges of incorporating prior information into FWI. \Cref{sec:methodology} presents the proposed PnP-Flow formulation and reconstruction algorithm. \Cref{sec:experiments} describes the experimental setup and evaluates the proposed method first on two families of circular-inclusion models and then under distribution shift using breast phantom images. \Cref{sec:discussion} discusses the main findings and limitations of the proposed approach.
Appendix provides details of the optimization algorithm, the training data and learned priors, and the sensitivity of reconstruction to measurement noise.

\section{Problem Formulation and Background}\label{sec:background}

\subsection{Full-waveform inversion}\label{sec:fwi_background}

Let $\Omega\subset\mathbb{R}^{d}$ denote the physical domain and let $m(\mathbf{x})>0$ denote the acoustic wave speed.
For the $s$-th source, the corresponding wavefield $u_s(\mathbf{x},{t_{\mathrm w}};m)$ satisfies the wave equation,
\begin{equation}\label{eq:acoustic_wave_equation}
\frac{1}{m^2(\mathbf{x})}\frac{\partial^2 u_s}{\partial {t}_{{\mathrm w}}^2} - \Delta u_s = q_s(\mathbf{x},{t_{\mathrm w}}), \qquad (\mathbf{x},{t_{\mathrm w}})\in\Omega\times(0,T],
\end{equation}
together with appropriate initial and boundary conditions. {Here $t_{\mathrm w}$ denotes physical wave time;} and $q_s(\mathbf{x},{t_{\mathrm w}})$ denotes the source term\footnote{For simplicity, we use an interior source in this general formulation for illustrative purposes.
Depending on the application, the source may instead be prescribed on the boundary.
In our numerical experiments, both the sources and receivers are located on the boundary; the specific acquisition configuration used is described in Section \ref{sec:experiments}.}.
Let $\mathcal{P}$ denote the observation operator that samples the wavefield at the detector locations.
The measured data associated with source $s$ are modeled as
\begin{equation}\label{eq:fwi_single_source_data}
\mathbf{d}_s^\delta = \mathcal{P}u_s(m) + \boldsymbol{\eta}_s,
\end{equation}
where $\boldsymbol{\eta}_s$ represents measurement noise and modeling error, and $\delta=\|\boldsymbol{\eta}_s\|$ represents the noise level. We also write noiseless data directly as $\mathbf{d}_s = \mathcal{P}u_s(m)$.
We define the source-dependent forward operator by
\begin{equation}\label{eq:fwi_source_forward}
\mathcal{F}_s(m) := \mathcal{P}u_s(m).
\end{equation}
Collecting the measurements from all $N_s$ sources gives
\begin{equation}\label{eq:fwi_stacked_forward}
\mathcal{F}(m) := \begin{bmatrix}
\mathcal{F}_1(m)\\
\vdots\\
\mathcal{F}_{N_s}(m)
\end{bmatrix}, \qquad
\mathbf{d}^\delta := \begin{bmatrix}
\mathbf{d}_1^\delta\\
\vdots\\
\mathbf{d}_{N_s}^\delta
\end{bmatrix}.
\end{equation}
FWI amounts to inverting the nonlinear mapping $\mathcal{F}:\mathcal{X}\longrightarrow\mathcal{Y}$, where $\mathcal{X}$ denotes the set of admissible wave-speed fields and $\mathcal{Y}:=\prod_{s=1}^{N_s}\mathcal{Y}_s$ is the data space containing the detector-time traces associated with all source excitations.
For example, if the wavefield generated by each source is recorded at $N_r$ detector locations over the time interval $(0,T)$, one may take $\mathcal{Y}_s = L^2(0,T;\mathbb{R}^{N_r})$, together with the corresponding product norm on $\mathcal{Y}$.
In the discretized setting, $\mathcal{X}$ and $\mathcal{Y}$ reduce to finite-dimensional representations of the wave-speed field and the measured waveform data, respectively.

Since the nonlinear forward map does not generally admit an explicit inversion formula, FWI is commonly formulated as a numerical optimization problem.
A standard approach that estimates the wave-speed field by minimizing the squared $L^2$ discrepancy between the predicted and observed waveforms:
\begin{equation}\label{eq:fwi_optimization}
m^\star = \arg\min_{m\in\mathcal{X}} \Phi(m;\mathbf{d}^\delta),
\end{equation}
where
\begin{equation}\label{eq:fwi_data_fidelity}
\Phi(m;\mathbf{d}^\delta) := \frac{1}{2}\left\|\mathcal{F}(m)-\mathbf{d}^\delta
\right\|_{\mathcal{Y}}^2 = \frac{1}{2}\sum_{s=1}^{N_s}\left\|
\mathcal{P}u_s(m)-\mathbf{d}_s^\delta\right\|_{L^2(0,T;\mathbb{R}^{N_r})}^2.
\end{equation}

\subsection{Regularization and prior modeling in FWI} \label{sec:regularization_background}

The nonlinear least-squares problem \eqref{eq:fwi_optimization} arising from FWI has been studied extensively over the past several decades (see, e.g., \cite{tarantola1984inversion,pratt1999seismic,virieux2010overview,guasch2020full}).
Its gradient can be evaluated efficiently using the adjoint-state method \cite{plessix2006review,bates2022probabilistic}, which avoids explicitly forming the full Jacobian of the parameter-to-data map. Nevertheless, each gradient evaluation requires forward and adjoint wave-equation solves, typically for multiple source excitations. Consequently, repeated optimization at high spatial and temporal resolutions can remain computationally demanding.

As discussed in \Cref{sec:intro}, the waveform-misfit objective is nonlinear and generally nonconvex, while practical measurements may leave some components of the wave-speed field only weakly constrained. Consequently, minimizing the data-misfit term alone may not yield a stable or physically plausible reconstruction. A standard approach is therefore to incorporate prior information about the unknown wave-speed field through regularization, leading to
\begin{equation}\label{eq:regularized_fwi}
m^\star = \arg\min_{m\in\mathcal{X}} \left\{\Phi(m;\mathbf{d}^\delta)+\lambda\mathcal{R}(m)\right\},
\end{equation}
where $\mathcal{R}$ encodes prior assumptions about the admissible wave-speed field and $\lambda>0$ controls the balance between data fidelity and regularization. Classical examples include Tikhonov-type regularization,
\begin{equation}\label{eq:tikhonov_fwi}
\mathcal{R}_{\mathrm{Tik}}(m)=\|\mathcal{L}m\|_{L^2(\Omega)}^2,
\end{equation}
where $\mathcal{L}$ may represent the identity or a spatial derivative operator, and the total variation regularization,
\begin{equation}\label{eq:tv_fwi}
\mathcal{R}_{\mathrm{TV}}(m)=\|\nabla m\|_{L^1(\Omega)}.
\end{equation}
After spatial discretization, these regularizers give rise to familiar $\ell^2$- and $\ell^1$-type penalties.
In particular, when $\mathcal{L}$ is a derivative operator, Tikhonov regularization \eqref{eq:tikhonov_fwi} favors spatially smooth wave-speed fields by penalizing large spatial variations.
In contrast, total variation \eqref{eq:tv_fwi} promotes sparsity of the spatial gradient and is therefore well suited to preserving sharp interfaces and piecewise-smooth structures. We refer interested readers to \cite{modrak2016seismic,esser2018total} and the references therein for further discussion of these classical regularization approaches.

As motivated in \Cref{sec:intro}, an alternative to prescribing prior information through a fixed analytic penalty is to learn it directly from representative wave-speed fields; see, for example, \cite{arridge2019solving,lunz2018adversarial}. In this work, we use a generative model to learn statistical information about the distribution of plausible wave-speed fields and investigate how such a learned model can be incorporated into FWI as a flexible source of prior information.

This perspective is fundamentally different from learning an inverse map directly from measurements to wave-speed fields. A supervised inversion network seeks to approximate a mapping from waveform data to the corresponding model and therefore typically requires paired measurement-model training examples. Such a learned inverse map may also depend on the acquisition geometry, source configuration, and forward model used to generate the training data. In contrast, a generative prior can be trained using wave-speed fields alone, thereby separating the learning of model statistics from the measurement process. Once trained, the prior can be coupled with observed data through the physics-based forward model used in the inversion. Importantly, such a learned prior need not be represented by an explicit regularization functional $\mathcal{R}(m)$. Instead, its information can be introduced through an operator derived from the generative model, while the wave-equation data-fidelity term continues to enforce consistency with the observations. This separation naturally leads to a plug-and-play formulation, in which the learned model acts as a prior operator within an iterative physics-based reconstruction procedure. In the present work, we realize this idea using a flow-matching generative model and incorporate the resulting flow-based operator into FWI through this plug-and-play mechanism; see \Cref{sec:methodology} for details.

\section{Proposed Methodology}\label{sec:methodology}

\subsection{PnP-Flow formulation for FWI}\label{sec:pnp_flow_fwi_overview}

In this work, we adapt the PnP-Flow approach of \cite{martin2025pnpflow} to the FWI problem described in \Cref{sec:fwi_background}, coupling a learned flow-matching denoiser with PDE-constrained data-consistency updates. To begin with, we review the construction of the denoising operator from a flow-matching generative model following \cite{martin2025pnpflow}.
Specifically, let $P_0$ denote a simple latent distribution ($\mathcal{N}(0,I)$ in our case) and $P_1$ denote the target distribution of admissible model parameters (e.g., wave field with several circles inside).
A time-dependent vector field $v_\theta:\mathbb{R}^{d_m}\times[0,1]\to\mathbb{R}^{d_m}$ is trained through the conditional flow-matching objective
\begin{equation}\label{eq:cfm_loss}
	\mathcal{L}_{\mathrm{CFM}}(\theta)=
	\mathbb{E}_{\substack\{t\sim\mathcal{U}[0,1], z_0\sim P_0,z_1\sim P_1\}} \left[\left\| v_\theta((1-t)z_0+tz_1,t) - (z_1-z_0) \right\|_2^2\right].
\end{equation}
The corresponding learned denoiser is
\begin{equation}\label{eq:D-denoiser}
	D_t^{\theta}(z)=z+(1-t)v_{\theta}(z,t) .
\end{equation}
For the ideal vector field minimizing the population flow-matching loss, the corresponding operator equals the conditional mean of the clean target $z_1$ given the interpolated input $(1-t)z_0+tz_1=z$.
The learned operator $D_t^\theta$ approximates this estimator, motivating its use as a denoiser.

We then can incorporate this learned prior information $v_\theta$ into the FWI formulation of \Cref{sec:fwi_background} using the PnP-Flow construction of \cite{martin2025pnpflow}.
Let $\mathbf{m}\in\mathbb{R}^{d_m}$ denote the discrete representation of the wave-speed function $m(\mathbf{x})$.
Below, we outline the PnP-Flow formulation for FWI, which alternates between PDE-constrained data-consistency updates and learned denoising updates.

Let $\mathbf{m}_n$ denote the current wave-speed estimate at the $n$-th outer iteration.
We first perform a data-consistency update using only the FWI objective.
To keep the formulation independent of the particular optimization method, we write
\begin{equation}\label{eq:pnp_flow_fwi_data_step}
\mathbf{z}_n= \DataUpdate
\big(\mathbf{m}_n;\mathbf{d}^{\delta}\big),
\end{equation}
where $\DataUpdate$ denotes one or more optimization steps applied to the data-fidelity term $\Phi(\mathbf{m};\mathbf{d}^{\delta})$.
For example, a single gradient-descent step corresponds to $\mathbf{z}_n=\mathbf{m}_n-\eta_n\nabla\Phi(\mathbf{m}_n;\mathbf{d}^{\delta}), \eta_n>0$ as in \cite{martin2025pnpflow}, whereas in our formulation, $\DataUpdate$ consists of several Adam updates before the learned-prior step is applied.
In either case, this part of the iteration depends only on the observed waveform data and the wave-equation forward model; the learned prior does not enter the data-consistency block.

 The output $\mathbf{z}_n$ is then coupled with the learned flow prior. The flow-based denoiser at time $t$ is designed to act on states associated with the corresponding intermediate level of the flow path. Since $\mathbf{z}_n$ is produced by the FWI update and does not generally have this form, PnP-Flow introduces an intermediate projection step. Drawing an independent sample $\boldsymbol{\xi}_n\sim\mathcal{N}(0,I)$ from the base distribution, we form
 \begin{equation}\label{eq:pnp_flow_fwi_interpolation}
 \widetilde{\mathbf{z}}_n=(1-t_n)\boldsymbol{\xi}_n+t_n\mathbf{z}_n.
 \end{equation}
 This interpolation embeds the current data-consistent estimate at the prescribed flow level and produces the input on which the time-dependent prior operator is applied.

 The flow-based denoising update is then given by applying the learned denoiser \eqref{eq:D-denoiser},
 \begin{equation}\label{eq:pnp_flow_fwi_denoising}
 \mathbf{m}_{n+1}=D_{t_n}^{\theta}(\widetilde{\mathbf{z}}_n)=
 \widetilde{\mathbf{z}}_n+(1-t_n)v_{\theta}(
 \widetilde{\mathbf{z}}_n,t_n).
 \end{equation}
The operator $D_{t_n}^{\theta}$ incorporates statistical information learned from the training ensemble of wave-speed models and moves the current estimate toward models favored by that learned distribution.
Following the PnP-Flow construction~\cite{martin2025pnpflow}, we form multiple interpolated fields using independent noise samples in \eqref{eq:pnp_flow_fwi_interpolation} and average their denoised outputs as in \eqref{eq:pnp_flow_fwi_denoising}, with the aim of improving the robustness of each prior update. This can be written as $\mathbf{m}_{n+1}=\sum_{l=1}^M D_{t_n}^{\theta}((1-t_n)\boldsymbol{\xi}_{n,l}+t_n\mathbf{z}_n) /M$.

An outer PnP-Flow-FWI iteration therefore alternates between two distinct sources of information, with the reprojection step mediating between them.
The data-consistency block \eqref{eq:pnp_flow_fwi_data_step} is governed entirely by the measured waveforms and the wave-equation physics, whereas the denoising update \eqref{eq:pnp_flow_fwi_denoising} incorporates information learned from representative wave-speed models.
This modular structure is particularly useful for FWI: the flow model can be trained offline using wave-speed models alone, without paired waveform--model examples, while the acquisition geometry and observed data enter only through the FWI objective during reconstruction.
Consequently, the learned prior does not explicitly encode a particular source--receiver configuration and can, in principle, be reused with different measurements associated with the same class of wave-speed models.

Additionally, the flow-time parameter $t_n$ controls the interaction between these two components.
Through \eqref{eq:pnp_flow_fwi_interpolation}, it determines how strongly the current data-consistent estimate is mixed with a sample from the base distribution, while through \eqref{eq:pnp_flow_fwi_denoising}, it also controls the magnitude and character of the subsequent flow-based prior update.
A prescribed schedule for $t_n$ therefore provides a natural mechanism for varying the influence of the learned prior over the course of the reconstruction.

Equations \eqref{eq:pnp_flow_fwi_data_step}, \eqref{eq:pnp_flow_fwi_interpolation}, and \eqref{eq:pnp_flow_fwi_denoising} define the proposed PnP-Flow FWI reconstruction, with details summarized in \Cref{sec:algorithm}.

\begin{remark}
	To relate the PnP-Flow to classical regularization, forward--backward splitting applied to the spatial discretization of \eqref{eq:regularized_fwi} would follow the single gradient-descent step in \eqref{eq:pnp_flow_fwi_data_step} with the proximal update
	\begin{equation}\label{eq:fwi_proximal_step}
		\mathbf{m}_{n+1}=\operatorname{prox}_{\eta_n\mathcal{R}}(\mathbf{z}_n),
	\end{equation}
where $\mathcal{R}$ here denotes the discretized regularization functional, with any admissibility constraints included through an indicator function.
		The central idea of PnP is to replace this proximal step by a denoising operator, thereby incorporating prior information without requiring an explicit functional $\mathcal{R}$~\cite{venkatakrishnan-2013-pnp}.
		For the proposed FWI method, this role is played by a flow-based denoiser.
\end{remark}

\subsection{Algorithms}\label{sec:algorithm}

 We now summarize our algorithm discussed in \Cref{sec:pnp_flow_fwi_overview},
 which alternates
 data-consistency updates with a learned flow-matching denoiser.
 The algorithm performs multiple Adam iterations \cite{kingma2015adam} between
 successive denoiser/prior updates, to fully enforce physics and data before restricting the reconstruction to the pretrained denoiser distribution.  The complete procedure is
 presented in \Cref{alg:pnp-fm-adam}.

\begin{algorithm}[h]
\caption{PnP-Flow full-waveform inversion}
\label{alg:pnp-fm-adam}
\begin{algorithmic}[1]
\State \textbf{Input:} Initial wave-speed model $\mathbf{m}_0$, data-misfit objective
$\Phi(\,\cdot\,;\mathbf{d}^\delta)$ and model gradient
$\nabla\Phi(\,\cdot\,;\mathbf{d}^\delta)$,
pretrained flow-matching vector field $v_\theta$,
flow starting time $\tstart$ and cap time $\tmax$,
numbers of outer iterations $\Nout$, inner iterations $\Nin$, and samples inside denoiser $M$,
base learning rate $\eta_0$, affine transform and reconstruction map
$\A,\A^{-1}$, and Adam parameters in \Cref{alg:adam-direction}
\State $\Delta t=(\tmax-\tstart)/(\Nout\Nin)$, $t_{-1}=\tstart$
\For{$n=0,\ldots,\Nout-1$}
    \State \textbf{\emph{Inner optimization iteration:}}
    \State $\mathbf{m}^{(0)}=\mathbf{m}_n$, $t^{(0)}=t_{n-1}$; reset both Adam moments to zero
    \For{$k=0,\ldots,\Nin-1$}
        \State $\mathbf{g}^{(k)}=\nabla\Phi(\mathbf{m}^{(k)};\mathbf{d}^\delta)$
        \State $\eta^{(k)}=\eta_0(1-t^{(k)})$
        \State Compute $\dmAdam^{(k)}(\mathbf{g}^{(k)})$ and update the Adam moments using \Cref{alg:adam-direction}
        \State $\mathbf{m}^{(k+1)}=\mathbf{m}^{(k)}+\eta^{(k)}\dmAdam^{(k)}(\mathbf{g}^{(k)})$
        \State $t^{(k+1)}=t^{(k)}+\Delta t$
    \EndFor
    \State $t_n=t^{(\Nin)}$
    \State \textbf{\emph{Apply flow matching denoiser:}}
    \State $\mathbf{z}_n=\A(\mathbf{m}^{(\Nin)})$
    \Comment{affine transform for normalization}
    \For{$\ell=1,\ldots,M$}
        \State Draw an independent Gaussian noise sample $\boldsymbol{\xi}_{n,\ell}\sim\mathcal{N}(0,I)$
        \State $\widetilde{\mathbf{z}}_{n,\ell}=(1-t_n)\boldsymbol{\xi}_{n,\ell}+t_n\mathbf{z}_n$
        \Comment{linear interpolation}
        \State $\widetilde{\mathbf{m}}_{n+1,\ell}
        =D_{t_n}^{\theta}(\widetilde{\mathbf{z}}_{n,\ell})
        =\widetilde{\mathbf{z}}_{n,\ell}
        +(1-t_n)v_\theta(\widetilde{\mathbf{z}}_{n,\ell},t_n)$ \Comment{apply FM denoiser}
    \EndFor
    \State $\mathbf{m}_{n+1}=\displaystyle\frac{1}{M}\sum_{\ell=1}^{M}
    \A^{-1}(\widetilde{\mathbf{m}}_{n+1,\ell})$
    \Comment{average reconstructed proposals}
\EndFor
\State \Return Reconstructed wave-speed model $\mathbf{m}_{\Nout}$
\end{algorithmic}
\end{algorithm}

 \begin{remark}
  In \Cref{alg:pnp-fm-adam}, the Adam states are reset at the beginning of each outer
 	iteration.

     The affine map
 	$\A$ combines the periodic grid embedding with the normalization from physical
 	wave-speed units to the prior-model range of $P_1$.  Its inverse $\A^{-1}$ denormalizes
 	the flow output and removes the duplicated periodic endpoint, thereby
 	recovering the field on the original PDE grid of wave-speed field.
 \end{remark}

\begin{remark}[Convergence]
Martin et al.~\cite{martin2025pnpflow} establish convergence of the original PnP-Flow iteration under the assumptions of their Proposition 4.
Our method adapts this iteration to nonlinear waveform inversion using multiple Adam updates between prior updates.
A convergence analysis of this modified iteration remains open.
\end{remark}

The two components in \Cref{alg:pnp-fm-adam}, Adam block and FM denoiser, play complementary roles.
The Adam iterations promote
consistency with the measured data and the governing wave equation, whereas
the flow-matching update transports the reconstruction toward the learned
training-data distribution.
A smaller $\tstart$ assigns more weight to Gaussian noise and produces a longer flow
trajectory, resulting in stronger prior regularization.

The computational cost of each Adam iteration in \Cref{alg:pnp-fm-adam} is dominated by evaluating the
multi-source objective and its gradient.
Specifically, $\Phi(\mathbf{m};\mathbf{d}^\delta)$ accumulates the misfit over $N_s$ sources, each involving a solution $u_s$ to the wave equation. Evaluating the misfit for each source requires one forward wave-equation solve, while computing its gradient requires one extra adjoint equation solve.
Consequently, each Adam iteration requires
approximately $2\Ns$ PDE solves, and one data-consistency block requires approximately $2\Ns\Nin$ PDE solves. In comparison, each flow-matching prior update requires $M$ evaluations of the learned vector field $v_\theta$.
The parameters $\Nin$ and $M$ determine the computational effort assigned to the data-consistency and the prior updates,
respectively, while $\tstart$ controls the strength of the prior transport. Their
selection therefore balances computational cost, physical consistency, and
learned regularization.

\section{Numerical Experiments}\label{sec:experiments}
We ask whether a flow prior trained on sound-speed fields containing inclusions improves reconstruction compared with TV-FWI when both methods use the same number of waveform-misfit updates per run, how it affects recovered inclusion contrast, and whether any improvements extend to breast phantoms without retraining or retuning.
We focus on small inclusions and examine how regularization affects their recovered contrast.
We first assess reconstruction accuracy and contrast recovery for circular inclusions in constant and spatially varying backgrounds, using priors trained in their respective families (\Cref{sec:exp-flat,sec:exp-gauss}).
To examine whether the benefits extend beyond these training families, we then apply the Gaussian-circle prior to breast phantoms without retraining and retain each method's reconstruction settings from the Gaussian-circle experiments
without retuning (\Cref{sec:exp-breast}).

\subsection{Experimental setup}\label{sec:exp-protocol}

\begin{figure}[tbp]
\centering
\includegraphics[width=0.8\linewidth]{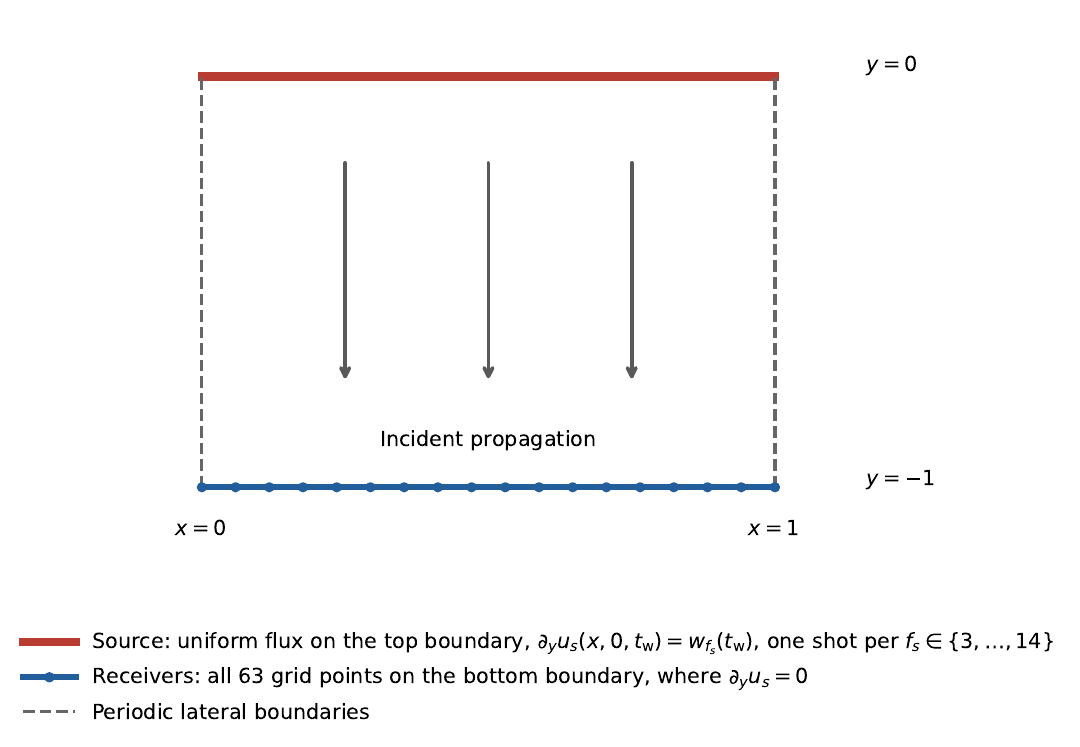}
\caption{Plane-wave acquisition shared by the three studies.
A spatially uniform source flux is applied at the top boundary in each frequency-specific shot, with waves propagating toward the bottom receivers.
}
\label{fig:acquisition}\label{fig:setup}
\end{figure}

\medskip
\noindent\textbf{Forward model and data acquisition.}
In our numerical experiments, we consider FWI with a forward model that follows the acoustic wave equation \eqref{eq:acoustic_wave_equation} on the domain \(\Omega=(0,1)\times(-1,0)\), with zero initial conditions, periodic boundary conditions on the left and right sides, and homogeneous Neumann boundary conditions on the bottom.
The spatially uniform sources are imposed through the top-boundary flux \(\partial_y u_s(x,0,t_{\mathrm w})=w_{f_s}(t_{\mathrm w})\) for each shot \(s\), which follow a Ricker time profile and differ in their central frequencies \(f_s\) (illustrated in \Cref{fig:setup}),
\[
w_{f_s}(t_{\mathrm w}):=\left[1-2\pi^2f_s^2(t_{\mathrm w}-1.5/f_s)^2\right]
\exp[-\pi^2f_s^2(t_{\mathrm w}-1.5/f_s)^2].
\]
There is no internal source in our experiments, i.e., $q_s = 0$ in \eqref{eq:acoustic_wave_equation}.
In total, we use twelve shots with central frequencies \(f_s\in\{3,4,\ldots,14\}\), and the receivers are all placed at the bottom boundary of the domain.

The wave equation is discretized using centered second-order finite differences in space and an explicit second-order time-stepping scheme. The spatial grid contains \(63\) horizontal and \(64\) vertical grid points.
For each target wave-speed field \(\mathbf m^\dagger\), the time step is chosen as \(\Delta t={0.4}/({63\sqrt{2}\,\max \mathbf m^\dagger})\) to satisfy the CFL condition and is kept fixed throughout the corresponding inversion.
All \(63\) grid points $\{x_r\}_{r=1}^{63}$ on the bottom boundary are used as receivers.
We simulate \(1000\) time steps and record the wavefield at the receivers every second time step:
\[ d_{s,r,n}=u_s((x_r,-1),2n\Delta t;\mathbf m^\dagger),\qquad n=0,1,2,\ldots,499.\]
These are directly used as our synthetic measurements $\mathbf d_s$ for each shot in our numerical experiments, i.e., we use noiseless data\footnote{The primary experiments use noiseless data. We also tested the method with noisy measurements; see \Cref{tab:supp-noise}. Since the results are qualitatively similar to those in the noiseless setting, we report the noiseless results in the main text for simplicity of presentation.}.
Thus, with the total data \(\mathbf d=(\mathbf d_1,\dots,\mathbf d_{12})\), the corresponding discrete waveform-misfit objective is
\[\Phi(\mathbf m;\mathbf d)=\frac12\sum_{s=1}^{12}\sum_{r=1}^{63}\sum_{n=0}^{499}\left|u_s((x_r,-1),2n\Delta t;\mathbf m)-d_{s,r,n}\right|^2.\]

\medskip
\noindent \textbf{Reconstruction algorithms and error metrics.}
We compare our proposed PnP-Flow method \Cref{alg:pnp-fm-adam} with the standard TV regularized method for FWI, which instead solves the following optimization problem
\begin{equation}
	\argmin_{\mathbf{m}}\; \Phi(\mathbf{m};\mathbf{d})+\lambda_{\mathrm{TV}}
	\mathcal{R}_{\mathrm{TV},\varepsilon_{\mathrm{TV}}}(\mathbf{m}),
\end{equation}
\[
\mathcal{R}_{\mathrm{TV},\varepsilon_{\mathrm{TV}}}(\mathbf m)
:=
\sum_{i,j}
\sqrt{
(m_{i+1,j}-m_{ij})^2
+
(m_{i,j+1}-m_{ij})^2
+
\varepsilon_{\mathrm{TV}}
},
\qquad
\varepsilon_{\mathrm{TV}}=10^{-4}.
\]
This optimization is also solved using the Adam iterations with a constant step size and the update direction specified in \Cref{alg:adam-direction}, which we later denote as TV-FWI.

Both TV-FWI and PnP-Flow are initialized from the constant field \(\mathbf m_0=\mathbf 8\).
PnP-Flow follows \Cref{alg:pnp-fm-adam}, using an FM denoiser pretrained on images containing several circles (details in \Cref{supp:data-selection-cost}),
with initial learning rate $\eta_0 = 0.1$, flow starting time $\tstart=0.5$ and cap time $\tmax=1$.
In the experiments,
we use $N_{\mathrm{out}}=50$ outer iterations, each consisting of
$N_{\mathrm{in}}=50$ inner optimization iterations, followed by one
flow-prior update. We refer to each inner optimization iteration as
a \emph{data update}. During each flow-prior update, $M=8$ independent
Gaussian noise samples are used to generate eight denoised proposals,
which are averaged to obtain the starting field for the next outer
iteration. Thus, one complete PnP-Flow reconstruction involves
$50\times 50=2500$ data updates and $50$ flow-prior updates.
To fairly compare the PnP-Flow and TV-FWI, we also use 2500 total Adam updates for TV-FWI, thus they are under similar computational costs. On an NVIDIA RTX 6000 Ada GPU, reconstruction took about 58~s for PnP-Flow and 52~s for TV-FWI.

For a target \(\mathbf m_i^\dagger\) and a reconstruction \(  \widehat{\mathbf{m}}_i\), we measure reconstruction accuracy using the relative \(L^2\) error on the solver grid,
\begin{equation}\label{eq:relative_l2}
e_i(\widehat{\mathbf{m}}_i) = {\|\widehat{\mathbf{m}}_i-\mathbf m_i^\dagger\|_2}/{\|\mathbf m_i^\dagger\|_2}.
\end{equation}
However, note that since the relative \(L^2\) error is computed over the entire image, it may remain small even when the contrast of a localized inclusion is poorly recovered. We therefore introduce a second metric that measures inclusion contrast. Specifically, for each inclusion \(j\) in target \(i\), we compare the mean wave speed inside the inclusion with that of the nearby background. Let \(K_{ij}\) denote the interior of the true inclusion after removing a one-pixel boundary layer, and let \(R_{ij}\) denote the surrounding background region consisting of pixels lying \(2\)--\(5\) pixels from the inclusion. Pixels belonging to any true inclusion are excluded from \(R_{ij}\). These regions are determined from the ground-truth target and kept fixed when evaluating all reconstructions. For a discrete wave-speed field \(\mathbf m\), we define the inclusion contrast by
\[C_{ij}(\mathbf m)=
\operatorname{mean}_{K_{ij}}\mathbf m-\operatorname{mean}_{R_{ij}}\mathbf m.\]
Thus, \(C_{ij}(\mathbf m)\) is positive when the inclusion has a higher mean wave speed than its local background and negative when it has a lower mean wave speed.
The corresponding contrast error for target \(i\) is then defined as
\begin{equation}\label{eq:constrast_error}
E_{C,i}(\widehat{\mathbf m}_i)=\frac{1}{n_i}\sum_{j=1}^{n_i}\bigl|C_{ij}(\widehat{\mathbf m}_i)-C_{ij}(\mathbf m_i^\dagger)\bigr|,
\end{equation}
where \(n_i\) is the number of inclusions evaluated in target \(i\). For the circle data (\Cref{sec:exp-flat}--\Cref{sec:exp-gauss}), each spatially separate inclusion region in the ground-truth target is evaluated. For the breast data (\Cref{sec:exp-breast}), only the designated target lesion is evaluated, so \(n_i=1\); pixels belonging to other annotated lesions are excluded from the surrounding background region. The contrast error has the same units as the wave-speed field.

Finally, since FM denoiser introduces stochasticity into PnP-Flow, we perform four independent PnP-Flow reconstructions for each target. For both the relative \(L^2\) error \eqref{eq:relative_l2} and the contrast error \eqref{eq:constrast_error}, we use the median over the four runs as the error for that target. TV-FWI is deterministic and is therefore evaluated once per target. For each method, we report the median error over all targets. We also report the number of targets for which the PnP-Flow median error is lower than the corresponding TV-FWI error.

\subsection{Circles on a constant background}\label{sec:exp-flat}

We first test whether PnP-Flow improves accuracy at matched data-update budgets on piecewise-constant fields containing circular inclusions, regions whose sound speed differs from the surrounding background.
We evaluate the methods on 50 test images containing faster and slower inclusions embedded in a uniform background of sound speed $8$. The learned prior is trained on separate images from the same family, with varying inclusion sizes, positions, and sound speeds, so this experiment provides an in-distribution comparison.
The uniform regions and sharp interfaces are also compatible with TV regularization, making TV-FWI a natural baseline for this comparison.
Thus, we compare with TV-FWI
with regularization parameter \(\lambda_{\mathrm{TV}}=0.0003\) and step size 0.1 selected using a separate set of constant-background images.

\begin{table}[tbp]
\caption{\textbf{Constant-background circles, all 50 images.}
Relative \(L^2\) errors are percentages, summarized by the median over images; PnP-Flow uses each image's median error over four independent runs.
Each run uses the listed budget, and no fields are averaged.
TV-FWI uses \(\lambda_{\mathrm{TV}}=0.0003\) and step size 0.1.
Contrast error is the inclusion contrast error of \Cref{sec:exp-protocol}, in model sound-speed units, summarized by the median over images.
}
\label{tab:flat-main}
\setlength{\tabcolsep}{3pt}
\setlength{\pwtablewidth}{\dimexpr\linewidth-6\tabcolsep\relax}
\begin{tabular}{@{}>{\raggedright\arraybackslash}p{0.35\pwtablewidth}>{\centering\arraybackslash}p{0.19\pwtablewidth}>{\raggedleft\arraybackslash}p{0.24\pwtablewidth}>{\raggedleft\arraybackslash}p{0.22\pwtablewidth}@{}}
\toprule
Reconstruction & Updates per run & Median relative \(L^2\) error (\%) & Median contrast error \\
\midrule
Constant start, \(m=8\) & 0 & 6.57 & 2.016 \\
PnP-Flow, median run error & 2,500 & 1.92 & 0.051 \\
TV-FWI & 2,500 & 2.65 & 0.316 \\
\botrule
\end{tabular}
\end{table}

At the nominal budget of 2,500 data updates per run, the median relative \(L^2\) error is \(1.92\%\) for PnP-Flow and \(2.65\%\) for TV-FWI; see \Cref{tab:flat-main}.
This improvement is also reflected in the target-by-target comparison: PnP-Flow has lower relative \(L^2\) error on 39 of 50 images.
The median inclusion contrast error is \(0.051\) for PnP-Flow and \(0.316\) for TV-FWI, and PnP-Flow has lower contrast error on 48 of 50 images.
Sensitivity tests on noisy data are reported in \Cref{tab:supp-noise}; at the tested noise levels, both error measures change little for either method, and PnP-Flow retains lower median errors in both measures.

\begin{figure}[tbp]
\centering
\includegraphics[width=0.98\linewidth]{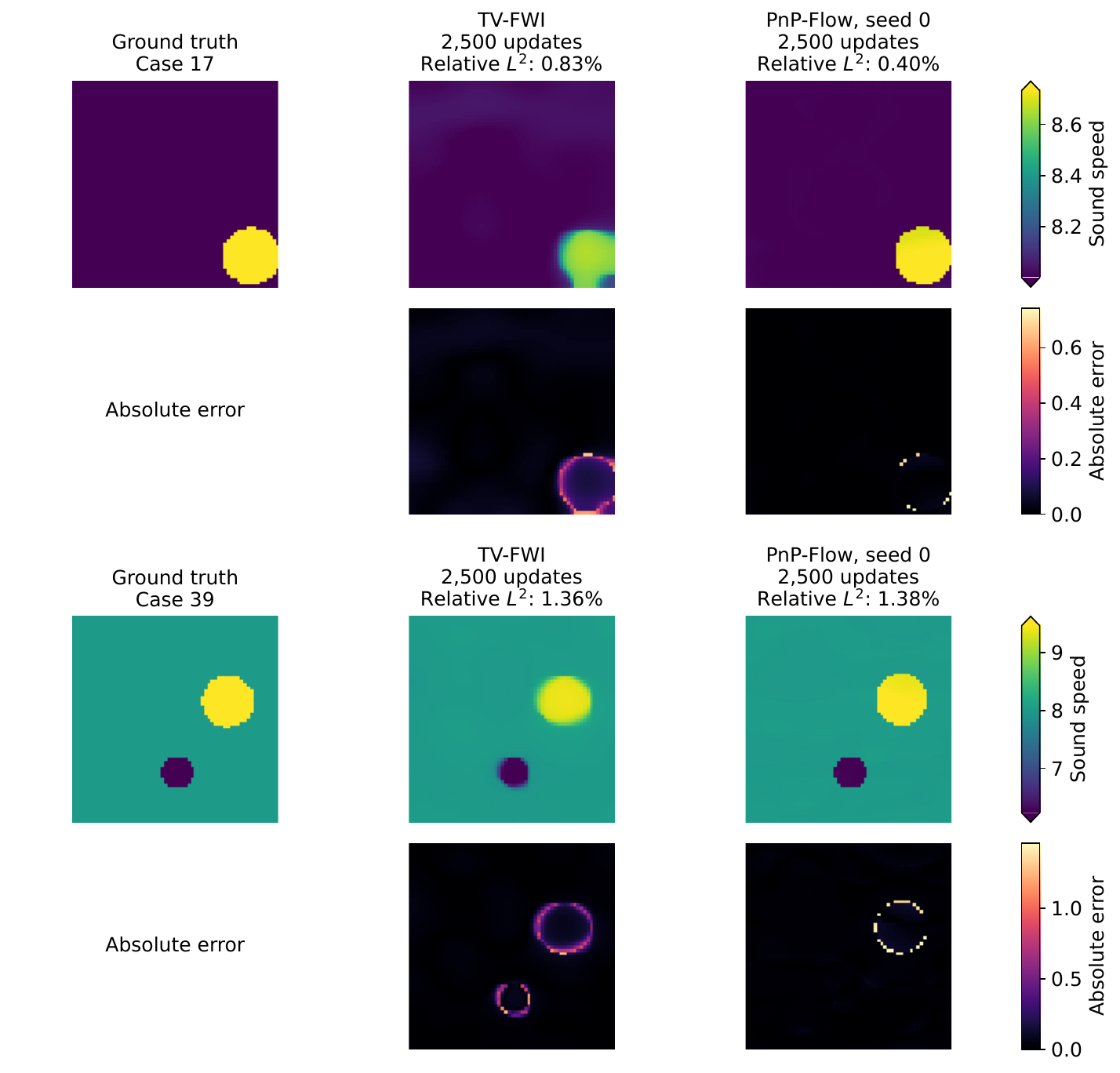}
\caption{%
Constant-background circles: reconstructions and absolute-error maps for TV-FWI and PnP-Flow (seed 0) after 2,500 updates. Cases 17 and 39 illustrate examples with lower relative \(L^2\) reconstruction errors for PnP-Flow and TV-FWI, respectively. Color scales are shared within each case.}

\label{fig:flat-panel}
\end{figure}
\Cref{fig:flat-panel} illustrates the reconstruction results for two representative test images. In Case 17, PnP-Flow recovers the inclusion with a sharper boundary and more accurate contrast, whereas TV-FWI produces a smoother boundary and underestimates the contrast. Consequently, PnP-Flow achieves a substantially lower relative \(L^2\) error (0.40\% versus 0.83\%). In Case 39, both methods recover the inclusion boundaries sharply, but PnP-Flow slightly displaces the boundary of the larger inclusion, resulting in a marginally higher reconstruction error than TV-FWI (1.38\% versus 1.36\%).

Thus PnP-Flow lowers reconstruction error relative to the selected TV-FWI configuration for most constant-background test images at the matched update budget.

\subsection{Circles on spatially varying Gaussian background}\label{sec:exp-gauss}

{We next consider a more heterogeneous target family in which the background wave speed varies spatially. We retain the same \(50\) inclusion geometries used in \Cref{sec:exp-flat}, but replace the constant background by spatially correlated Gaussian variations around the background value \(8\). These variations contain structure at multiple spatial scales, so that the reconstruction must recover both the localized inclusions and the spatially varying background. A separate flow-matching prior is trained on fields generated from this Gaussian-background family; this experiment therefore remains an in-distribution test. Details of the data generation and prior training are given in Appendix \ref{supp:data-selection-cost}}.

{TV-FWI uses \(\lambda_{\rm TV}=0.003\) and step size \(0.01\).
At the budget of \(2500\) data updates per reconstruction, the median relative \(L^2\) error over the \(50\) test targets is \(3.67\%\) for PnP-Flow and \(4.69\%\) for TV-FWI; see \Cref{tab:gauss-main}.
On a target-by-target basis, PnP-Flow has a smaller reconstruction error on \(48\) of the \(50\) targets.
Thus, under the reconstruction settings considered here, PnP-Flow yields lower relative \(L^2\) error than TV-FWI on most targets in the Gaussian-background test set.
In inclusion contrast, the median error is \(0.239\) for PnP-Flow and \(0.851\) for TV-FWI, and PnP-Flow has lower contrast error on 49 of 50 targets.
}

\begin{figure}[tbp]
\centering
\includegraphics[width=0.98\linewidth]{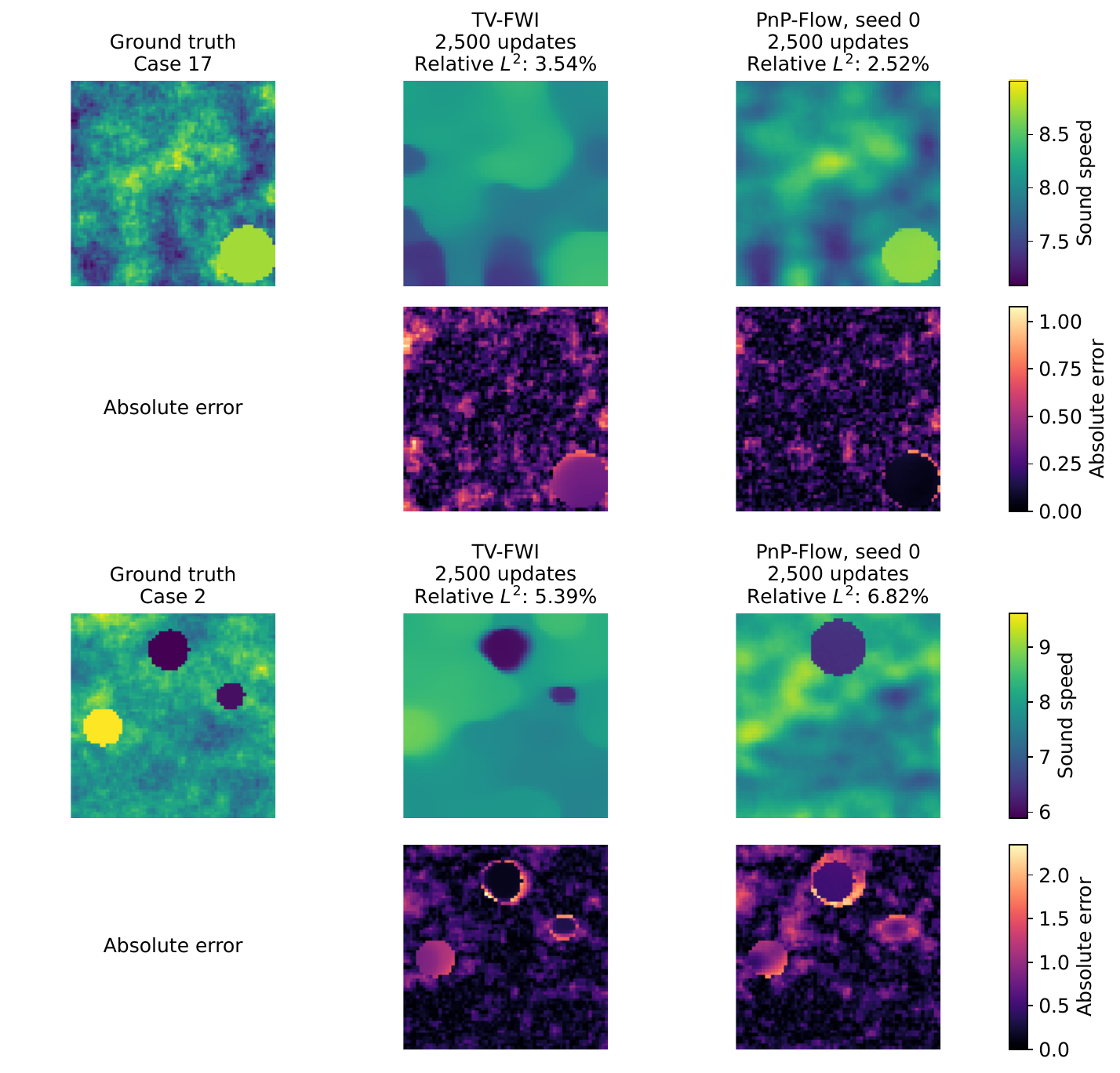}
\caption{%
Gaussian-background circles: reconstructions and corresponding absolute-error maps for TV-FWI and PnP-Flow (seed 0) after $2500$ data updates. Cases 17 and 2 yield lower relative \(L^2\) errors for PnP-Flow and TV-FWI, respectively. Case 17 uses the same inclusion geometry as case 17 in \Cref{fig:flat-panel}. Color scales are shared within each case.}
\label{fig:gauss-panel}
\end{figure}

Representative reconstructions are shown in \Cref{fig:gauss-panel}. Case 17 contains the same inclusion geometry as the corresponding constant-background example in \Cref{fig:flat-panel}. In this case, PnP-Flow recovers the inclusion boundary and most of the true contrast, whereas TV-FWI recovers only about half of the contrast. Case 2 is one of the two targets for which TV-FWI attains a lower relative \(L^2\) error. Here, PnP-Flow overestimates the extent of the large low-speed inclusion, and its recovered background is less accurate than the smoother background produced by TV-FWI. Neither method resolves the fine-scale background texture well, and both methods recover the high-speed inclusion only weakly in this example. These cases illustrate both the typical advantage of PnP-Flow in preserving inclusion contrast and a failure mode in which the learned prior distorts the inclusion geometry and background structure.

\begin{table}[tbp]
\caption{\textbf{Gaussian-background circles, all 50 images.} Entries are medians over images, and relative \(L^2\) errors are reported as percentages. For PnP-Flow, each image contributes the median error over four independent runs at the listed data-update budget; reconstructions are not averaged across runs. TV-FWI uses \(\lambda_{\mathrm{TV}}=0.003\) and step size \(0.01\), selected on a separate set of Gaussian background images.}
\label{tab:gauss-main}
\setlength{\tabcolsep}{3pt}
\setlength{\pwtablewidth}{\dimexpr\linewidth-6\tabcolsep\relax}
\begin{tabular}{@{}>{\raggedright\arraybackslash}p{0.35\pwtablewidth}>{\centering\arraybackslash}p{0.19\pwtablewidth}>{\raggedleft\arraybackslash}p{0.24\pwtablewidth}>{\raggedleft\arraybackslash}p{0.22\pwtablewidth}@{}}
\toprule
Reconstruction & Updates per run & Median relative \(L^2\) error (\%) & Median contrast error \\
\midrule
Constant start, \(m=8\) & 0 & 7.45 & 2.061 \\
PnP-Flow, median run error & 2,500 & 3.67 & 0.239 \\
TV-FWI & 2,500 & 4.69 & 0.851 \\
\botrule
\end{tabular}
\end{table}

\subsection{Lesion identification in breast phantom images}\label{sec:exp-breast}

{We conclude our comparisons with a more realistic scenario that resembles a medical imaging application. Specifically, we address the problem of finding lesions in breast phantom images. Similarly to the circles on spatially varying Gaussian background, the breast phantom images contain lesions that look like localized inclusions that must be identified inside a tissue with a varying texture. Although the interfaces and textures of the breast phantom case differ from the specific data of \Cref{sec:exp-gauss}, the setup is analogous: a non-homogeneous background and an inclusion identification task.
Hence, we take our trained model for the Gaussian-circle prior in the previous section without retraining and test how effective it is as a prior for the lesion-contrast reconstruction and detection in the breast phantom images case.
Note that given the somewhat limited availability of public breast imaging, in part due to privacy preservation reasons, it is useful to be able to build cheap data proxies to train sophisticated but data-hungry models, like diffusion generative models, and analyze performance under this intrinsic distribution shift. }

{For our experiments we utilize targets derived from the numerical acoustic breast phantoms of Li et al.~\cite{li20213}. These include four categories of breast tissue: `A' is predominantly fatty, `B' contains scattered fibroglandular tissue, `C' is heterogeneously dense, and `D' is extremely dense. We select 32 distinct \emph{parent} phantoms, eight per category, and build a test set containing 32 patches, each one extracted from a different phantom. Available phantom annotations allow us to select the target lesion and determine the patch size and location.}
{To apply our methodology to this acoustic data, sound speeds are mapped into model units by \(m=m_{\mathrm{m/s}}\,8/1450\), so that one model unit corresponds to approximately 181 m/s. New waveforms are generated using the acquisition described in \Cref{sec:exp-protocol}.

In the primary comparison, TV-FWI and PnP-Flow retain the parameter settings of \Cref{sec:exp-gauss}, without breast-specific tuning.
}%
We report relative \(L^2\) and lesion-contrast errors as defined in \Cref{sec:exp-protocol}, with the core and ring determined from phantom annotations.
\Cref{tab:breast-main} reports whole-image relative \(L^2\) and lesion-contrast errors for both methods at 2,500 updates, together with those of the constant starting field \(m=8\).
Both methods roughly halve the median whole-image error of the constant start, reducing it from \(2.99\%\) to \(1.56\%\) for PnP-Flow and \(1.52\%\) for TV-FWI.
Whole-image errors are similar, with PnP-Flow achieving lower error on 17 of the 32 patches and no clear overall advantage over TV-FWI.
In lesion contrast, however, PnP-Flow achieves a lower median error than TV-FWI, \(0.214\) versus \(0.263\) model sound-speed units, and has lower error on 28 of the 32 patches.

\begin{table}[tbp]
\caption{\textbf{Transfer to all 32 breast patches.}
TV-FWI uses the Gaussian-selected \(\lambda_{\mathrm{TV}}=0.003\) and step size 0.01; PnP-Flow uses the Gaussian-circle prior and reconstruction settings.
Entries are medians over patches; PnP-Flow uses each patch's median error over four independent runs without averaging fields.
Contrast errors are in model sound-speed units.
}
\label{tab:breast-main}
\setlength{\tabcolsep}{3pt}
\setlength{\pwtablewidth}{\dimexpr\linewidth-6\tabcolsep\relax}
\begin{tabular}{@{}>{\raggedright\arraybackslash}p{0.35\pwtablewidth}>{\centering\arraybackslash}p{0.19\pwtablewidth}>{\raggedleft\arraybackslash}p{0.24\pwtablewidth}>{\raggedleft\arraybackslash}p{0.22\pwtablewidth}@{}}
\toprule
Reconstruction & Updates per run & Median relative \(L^2\) error (\%) & Median contrast error \\
\midrule
Constant start, \(m=8\) & 0 & 2.99 & 0.381 \\
PnP-Flow, median run error & 2,500 & 1.56 & 0.214 \\
TV-FWI & 2,500 & 1.52 & 0.263 \\
\botrule
\end{tabular}
\end{table}

\begin{figure}[tbp]
\centering
\includegraphics[width=0.98\linewidth]{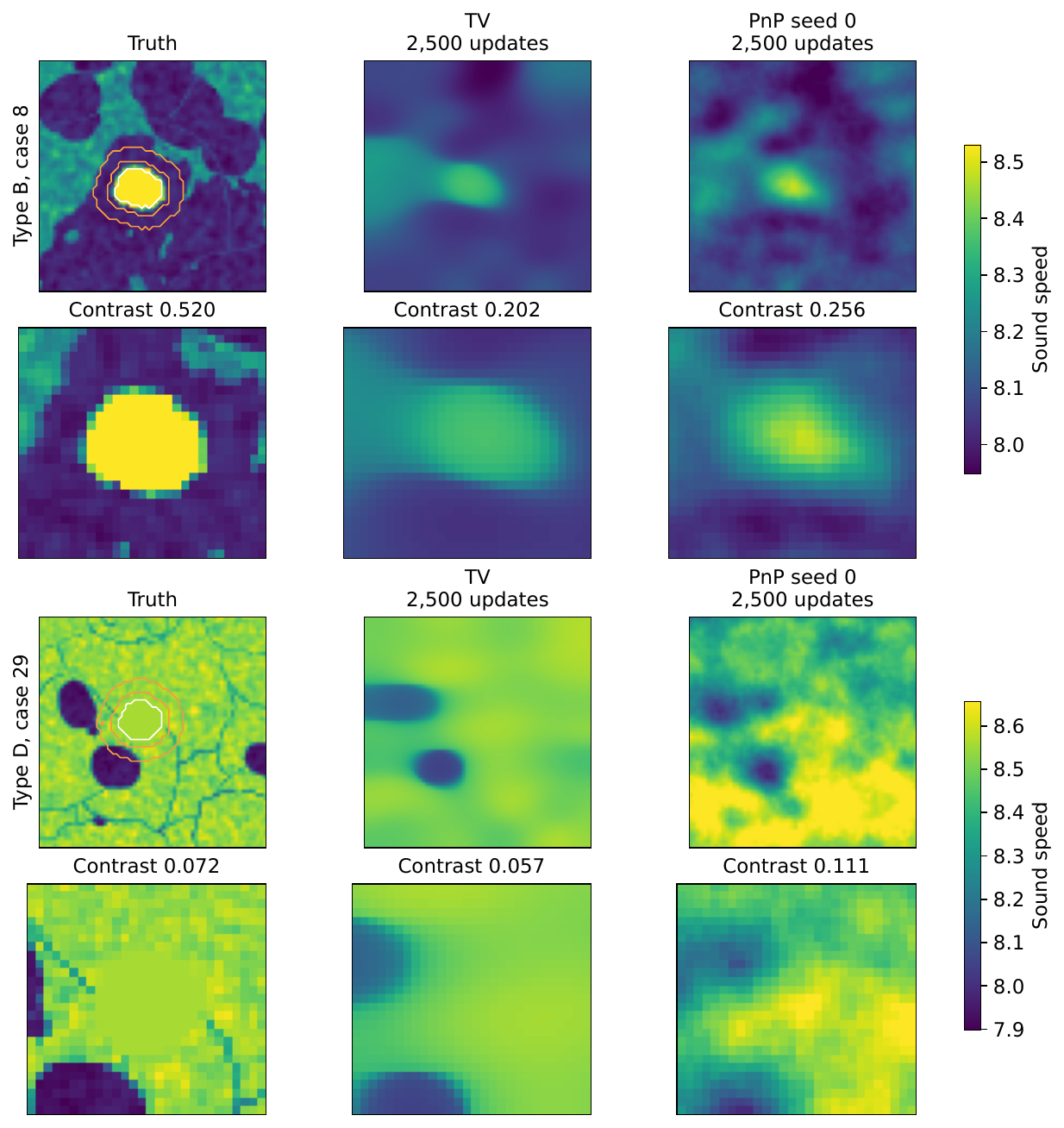}
\caption{Breast reconstructions and lesion-centered crops for TV-FWI and PnP-Flow seed 0, each after 2,500 updates, the budget of \Cref{tab:breast-main}.
Type-B case 8 and type-D case 29 favor PnP-Flow and TV-FWI in contrast error, respectively.
Scales are shared within each case; the truth panel marks the eroded lesion core in white and the ring in orange.}
\label{fig:breast-fields}
\end{figure}

PnP-Flow has lower contrast error on all eight patches of each of types A--C.
Results are mixed for type D: PnP-Flow has lower error on four of the eight patches, while TV-FWI has the lower median contrast error.
Type D also has the lowest median true lesion contrast among the four types in this cohort.
\Cref{fig:breast-fields} illustrates the variation across cases.
In the type-B example, both methods underestimate lesion contrast, but PnP-Flow recovers it more accurately.
In the type-D example, PnP-Flow overestimates the weaker lesion contrast, while TV-FWI is closer to the ground truth.

\section{Conclusion}
\label{sec:discussion}

We adapted PnP-Flow to nonlinear full-waveform inversion by alternating waveform-misfit updates with updates from a pretrained flow prior. At matched computational budgets, the proposed method achieves lower median whole-image and inclusion-contrast errors than TV-FWI on both circular-inclusion model families tested, i.e., circles on a constant background and circles on spatially varying Gaussian background. In both cases, the prior is learned from sound-speed images alone, while measurement information enters through the waveform-misfit updates during reconstruction. This allows separating the learning of model statistics from the measurement process, making for a general data-driven regularization methodology whose training does not depend on the acquisition geometry or source configuration.

In a more realistic scenario, the breast experiment shows that out-of-distribution performance is not assured and depends on the feature being evaluated.
With the Gaussian-circle prior and both methods' reconstruction settings transferred without breast-specific tuning, PnP-Flow improves lesion-contrast recovery while giving whole-image errors comparable to TV-FWI.
This distinction supports evaluating local contrast alongside global reconstruction error.
The PnP-Flow breast reconstructions also contain rounded structures resembling the training inclusions (\Cref{fig:breast-fields}), showing that lesion contrast can improve while tissue texture remains inaccurate.
For constant-background circles, PnP-Flow also
exhibits better performance than
TV-FWI across the tested measurement-noise levels (\Cref{tab:supp-noise}).

Future work directions include a complete theoretical convergence analysis of the PnP-Flow when solving nonlinear problems, and evaluation of transfer learning scenarios where synthetic data can be exploited to solve other experimental setups with limited data availability, as is the case for some medical imaging problems.

\backmatter

\bmhead{Acknowledgments}
The authors would like to acknowledge the support from the Research Collaboration Workshop in the Science of Data and Mathematics (WiSDM) hosted by the University of North Carolina at Chapel Hill from August 4-8, 2025, which initiated this work. F. Terzioglu's work was supported in part by NSF grant DMS-2206279. LZ is partially supported by the NSF grant DMS-2231482.

\section*{Declarations}
An AI-based language tool was used to assist in editing the manuscript for spelling, grammar, and stylistic improvements. The authors subsequently reviewed and revised the text as necessary and take full responsibility for the content of the published article.
The authors used Claude Code (Anthropic) to assist with writing and debugging the software used in the numerical experiments. All code was subsequently reviewed and tested by the authors, who take full responsibility for the software and the reported results.

\paragraph{Code availability.}
The code is available at \url{https://github.com/siting6/pnpflow-fwi-numerics}.

\begin{appendices}
\renewcommand{\theHfigure}{\thesection.\arabic{figure}}
\renewcommand{\theHtable}{\thesection.\arabic{table}}

\section{Adam optimization algorithm}\label{supp:Adam}
We summarize the clipped Adam optimization algorithm (without bias correction) used to update the physics-based directions derived from the PDE constraints and data misfit.

\begin{algorithm}[h]
\caption{Adam Update Direction with Gradient Clipping}

\label{alg:adam-direction}
\begin{algorithmic}[1]
\State \textbf{Input:} Gradient $\mathbf{g}^{(k)}$, previous Adam state
$\AdamFirst^{(k)}$, $\AdamSecond^{(k)}$ ($\AdamFirst^{(0)} = \mathbf{0}$, $\AdamSecond^{(0)} = \mathbf{0}$), Adam parameters $\beta_1$, $\beta_2$,
and $\epsilon$, and gradient-clipping threshold $G$
\State $\widehat{\mathbf{g}}^{(k)}=\mathbf{g}^{(k)}
\min\{1,\,G/(\lVert \mathbf{g}^{(k)}\rVert_2+10^{-12})\}$
\State $\AdamFirst^{(k+1)}=\beta_1\AdamFirst^{(k)}
+(1-\beta_1)\widehat{\mathbf{g}}^{(k)}$
\State $\AdamSecond^{(k+1)}=\beta_2\AdamSecond^{(k)}
+(1-\beta_2)(\widehat{\mathbf{g}}^{(k)})^2$
\State $\dmAdam^{(k)} (\mathbf{g}^{(k)})=-\AdamFirst^{(k+1)}
/(\sqrt{\AdamSecond^{(k+1)}}+\epsilon)$
\State \Return Direction $\dmAdam^{(k)}$ and updated Adam state
$\AdamFirst^{(k+1)}$, $\AdamSecond^{(k+1)}$
\end{algorithmic}
\end{algorithm}

\begin{remark}
	The global $\ell_2$ clipping in \Cref{alg:adam-direction} preserves the gradient direction while limiting its norm to $G$, thereby reducing instability caused by unusually large FWI gradients.
The square
	$(\widehat{\mathbf{g}}^{(k)})^2$ denotes the element-wise square rather than an inner
	product or matrix product, and the Adam moments are not bias-corrected.
    In our numerical experiments, we use $(\beta_1,\beta_2)=(0.9,0.999)$,
    and clipping threshold $G=10$.
\end{remark}

\section{Training data and learned priors}\label{supp:data-selection-cost}
\label{supp:prior}

\begin{figure}[t]
\centering
\includegraphics[width=0.9\linewidth]{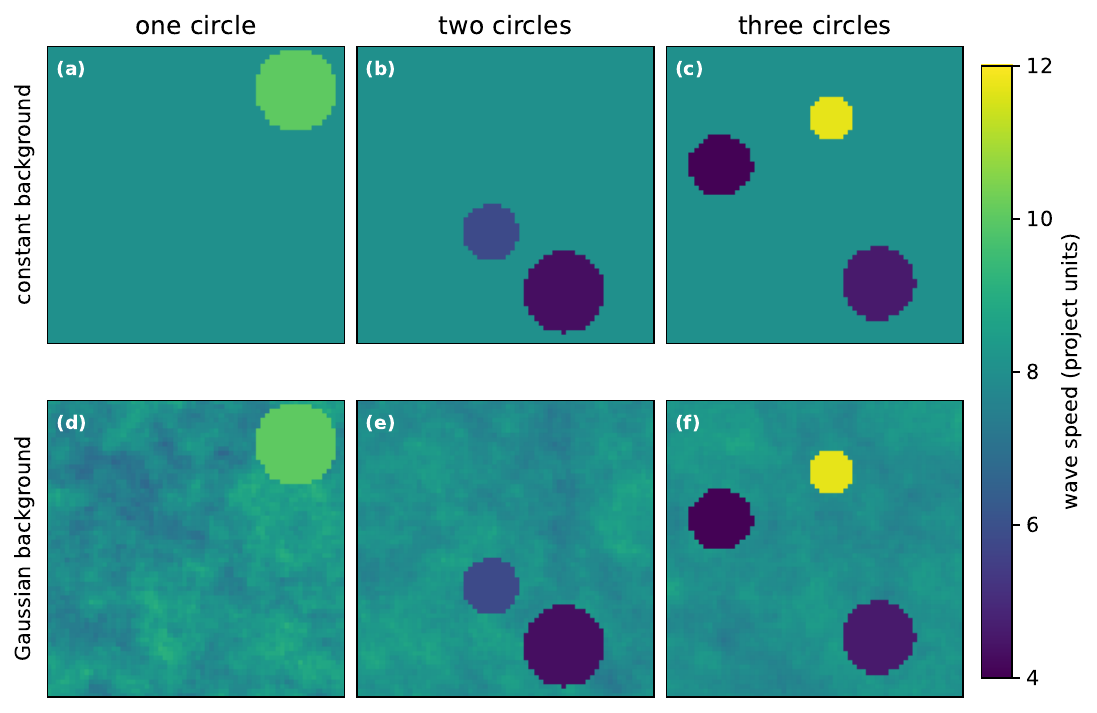}
\caption{Representative $64\times64$ training fields with constant
(top) and Gaussian (bottom) backgrounds and matching circle geometries.
Columns show one, two, and three inclusions.
}
\label{fig:prior-training-data}
\end{figure}

\noindent\textbf{Circle families.}
We construct two families of training fields that share the same circular inclusions but differ in their backgrounds.
Constant-background fields contain one to three nonoverlapping circular inclusions on a background of wave speed \(8\).
Inclusion wave speeds lie in \([4,12]\), and their radii range from \(0.06\) to \(0.14\) times the domain width.
We generate these fields on a \(256\times256\) grid and downsample them to \(64\times64\) using mode pooling.
From these downsampled fields, we construct the Gaussian-background family by retaining the inclusions and replacing the background with \(8+\sigma G\), where \(\sigma\sim\mathcal U[0.15,0.45]\).
Here, \(G\) is a standardized random field formed by combining Gaussian-smoothed noise at three spatial scales.
\Cref{fig:prior-training-data} illustrates the shared inclusion geometries and the two background types.
The two families share 4,884 training geometries, 100 validation geometries, and 50 test geometries held out from training and validation.
For the Gaussian family, independent background realizations are generated for the training, validation, and test sets.

\noindent\textbf{Prior training.}
Using these datasets, we normalize the training fields to \([-1,1]\) and train a separate prior for each family by optimal-transport conditional flow matching.
Both models use the same time-conditioned U-Net architecture and training configuration: 300 epochs, a batch size of 32, and a learning rate of \(2\times10^{-4}\) with warm-up followed by cosine decay.
During training, we maintain an exponential moving average of the model weights with decay \(0.9999\).
For each model, we select the checkpoint with the lowest validation loss.
The selected Gaussian-background prior is used for the Gaussian-circle experiments and reused for the breast-phantom transfer experiment.

\section{Sensitivity to measurement noise}\label{supp:noise}

\begin{table}[htbp!]
\centering
\caption{Sensitivity of the constant-background comparison to additive measurement noise.
We repeat the comparison of \Cref{tab:flat-main} on the same \(50\) test images, with measurement noise added to the data.
All reconstruction settings are kept at the values used for the noiseless data.
For each image and each shot, we add independent zero-mean Gaussian noise whose standard deviation is \(0.03\), \(0.10\) or \(0.316\) times the root-mean-square value of the clean shot record, which corresponds to \(30.5\)~dB, \(20\)~dB and \(10\)~dB; the same noise realization of each image is used for both methods.
We report median relative \(L^2\) and contrast errors at the nominal budget of \(2{,}500\) updates per run.
The PnP-Flow entries use four independent runs different from those of \Cref{tab:flat-main}, so its noiseless entries differ slightly from that table.}
\label{tab:supp-noise}
\setlength{\tabcolsep}{3pt}
\setlength{\pwtablewidth}{\dimexpr\linewidth-6\tabcolsep\relax}
\begin{tabular}{@{}>{\raggedright\arraybackslash}p{0.22\pwtablewidth}>{\raggedright\arraybackslash}p{0.30\pwtablewidth}>{\raggedleft\arraybackslash}p{0.24\pwtablewidth}>{\raggedleft\arraybackslash}p{0.24\pwtablewidth}@{}}
\toprule
Noise level & Reconstruction & Median relative \(L^2\) error (\%) & Median contrast error \\
\midrule
Noiseless & PnP-Flow & 1.79 & 0.049 \\
Noiseless & TV-FWI & 2.65 & 0.316 \\
30.5 dB & PnP-Flow & 1.77 & 0.048 \\
30.5 dB & TV-FWI & 2.66 & 0.312 \\
20 dB & PnP-Flow & 1.79 & 0.063 \\
20 dB & TV-FWI & 2.67 & 0.303 \\
10 dB & PnP-Flow & 1.89 & 0.063 \\
10 dB & TV-FWI & 2.84 & 0.319 \\
\botrule
\end{tabular}
\end{table}

In these sensitivity tests on noisy data, both methods show small changes in their median errors relative to the noiseless results, at most \(0.19\) percentage points in relative \(L^2\) error and \(0.014\) in contrast error.
Within this noise range, PnP-Flow has better performance than TV-FWI: at \(10\)~dB, it has lower relative \(L^2\) error on 42 of 50 images and lower contrast error on 49 of 50 images.

\end{appendices}

\bibliography{sn-bibliography}%

\end{document}